\documentclass[prd,superscriptaddress,a4paper,showpacs,showkeys,11pt,nofootinbib]{revtex4}
\usepackage{graphicx}
\usepackage{graphics}
\usepackage{dcolumn}
\usepackage{bm}
\usepackage{multirow}
\usepackage{tabularx}
\usepackage{hyperref}
\usepackage{commath}
\usepackage{epstopdf}
\usepackage[T1]{fontenc}
\usepackage[latin9]{inputenc}
\usepackage{geometry}
\usepackage{soul}
\usepackage{amsmath,amssymb,amsfonts}

\usepackage{color}

\def\be{\begin{equation}}
\def\ee{\end{equation}}

\providecommand{\ee}{e$^+$e$^-$}

\makeatother

\begin{document}

%\begin{flushright}
%MS-TP-22-36
%\end{flushright}

\title{Illuminating the charged leptons in the proton}

\author{Victor P. {\sc Gon\c{c}alves}}
\email{barros@ufpel.edu.br}
\affiliation{Institut f\"ur Theoretische Physik, Westf\"alische Wilhelms-Universit\"at M\"unster,
Wilhelm-Klemm-Stra\ss e 9, D-48149 M\"unster, Germany}
\affiliation{Institute of Modern Physics, Chinese Academy of Sciences,
  Lanzhou 730000, China}
\affiliation{Institute of Physics and Mathematics, Federal University of Pelotas, \\
  Postal Code 354,  96010-900, Pelotas, RS, Brazil}

\author{Daniel E. {\sc Martins}}
\email{daniel.ernani@ifj.edu.pl}
\affiliation{Institute of Physics and Mathematics, Federal University of Pelotas, \\
  Postal Code 354,  96010-900, Pelotas, RS, Brazil}
  \affiliation{The Henryk Niewodniczanski Institute of Nuclear Physics (IFJ) - Polish Academy of Sciences (PAN), \\
    31-342, Krakow, Poland}

\begin{abstract}
The  description of the structure of proton
is fundamental in order to describe the standard model processes at the LHC as well as for the searching of New Physics. Quantum fluctuations imply the presence of photons and leptons inside the proton, which admit a parton distribution function (PDF). Although the lepton PDFs  are expected to be small, its presence opens new production mechanisms. In order to explore the lepton - induced processes at the LHC, a precise determination of the leptonic content of the proton is needed. In this paper we propose to constrain  the content of charged leptons inside the proton through the study of the QED Compton scattering in ultraperipheral proton - nucleus collisions at the LHC. We estimate the total cross sections and associated distributions considering different models for the lepton PDFs and distinct lepton flavours. We demonstrate that a future experimental analysis of this process is feasible and that it can be used to constrain the content of electrons, muons and taus inside the proton. 
\end{abstract}

\keywords{Leptons, QED Compton Scattering, Ultraperipheral Collisions}

\maketitle

\section{Introduction}
One of the main goals of  Particle Physics is to achieve a deeper knowledge of the 
proton structure, which is fundamental to describe standard model (SM) processes and to identify possible signals of New Physics in hadronic collisions at the Large Hadron Collider (LHC) \cite{Gao:2017yyd,Kovarik:2019xvh,Amoroso:2022eow}. Our view of the proton structure has been largely improved with the successful operation of the DESY $ep$ collider HERA during the period between the years of 1992 and 2007, which has observed the striking rise of the proton structure function $F_2(x,Q^2)$ for small values of the Bjorken - $x$ variable ($\le 10^{-2}$) and a fixed photon virtuality $Q^2$. Such behaviour was interpreted in terms of the increasing  with the energy of the gluon and sea quark densities inside the proton \cite{Klein:2008di}. The advent of the LHC and the computation of higher - order QCD and electroweak (EW)  corrections for hadronic processes has motivated a huge progress in the determination of the parton distribution functions (PDFs) of the proton over the last years (For a recent review see, e.g. Ref. \cite{Amoroso:2022eow}). Recent results indicate the presence of an intrinsic charm component in the proton wave function \cite{Ball:2022qks}, as well as a photon and a lepton content in the proton (For recent results see, e.g. Refs. 
\cite{bertone,Bertone:2017bme,Harland-Lang:2019pla,buonocore,Cridge:2021pxm,Xie:2021equ}). In particular, the presence of leptons in the initial state of hadronic collisions opens the possibility of study lepton - lepton and lepton - quark subprocesses at the LHC \cite{Buonocore:2020erb,Harland-Lang:2021zvr,Buonocore:2021bsf}, with all combinations of charge and flavours, enlarging the scope of the LHC also for a lepton -- lepton (quark) collider. Such possibilities allow us to study new production channels, which can improve our understanding of the standard model and that can be used to search for Beyond - the - Standard - Model (BSM) Physics. However, in order to derive realistic predictions for these processes, a precise knowledge of the leptonic densities in the proton is fundamental.

Photons and leptons inside the proton can arise from quantum fluctuations, with photons being generated by the photon - quark splitting   process $q \rightarrow q \gamma$ and  charged leptons  by the photon splitting process $\gamma \rightarrow l^+ l^-$. In order to derive the photon and lepton PDF sets, one can implement such processes  in the Dokshitzer - Gribov - Altarelli - Lipatov - Parisi (DGLAP) evolution equations \cite{dglap} and perform a global fit of the existing data. However, such an alternative is still not viable due to the limited sensitivity of the LHC data for the photon and lepton initiated subprocesses. Another possibility is to compute these PDFs for a given scale on the base of a theoretically motivated model ansatz and derive the PDFs for other scales by solving the QED - corrected DGLAP equations,  as performed e.g. in Refs. \cite{Martin:2004dh} and \cite{bertone} for the photon and lepton cases, respectively (See also Refs. \cite{Martin:2004dh,Ball:2013hta,Bertone:2013vaa,Bertone:2016ume,Bertone:2017gds,Schmidt:2015zda}). Finally, a more precise determination of the photon and lepton PDFs in the proton can be performed using the LUX method, proposed originally for the photon case in Refs. \cite{lux1,lux2} and extended for charged leptons in Ref. \cite{buonocore}, where these PDFs are computed using only information from electron - proton scattering data. Such formalism was recently applied by the CTEQ - TEA \cite{Xie:2021equ} and MSHT \cite{Cridge:2021pxm} groups
 to derive new sets of parton distributions including the QED corrections to the DGLAP evolution.  Although such a method has improved  the accuracy of the predictions, the associated uncertainties are still non - negligible, which motivates the proposition of an alternative to probe the lepton PDFs in the proton using a physical process that can be measured at the LHC.  

\begin{figure}[t]
\includegraphics[scale=0.42]{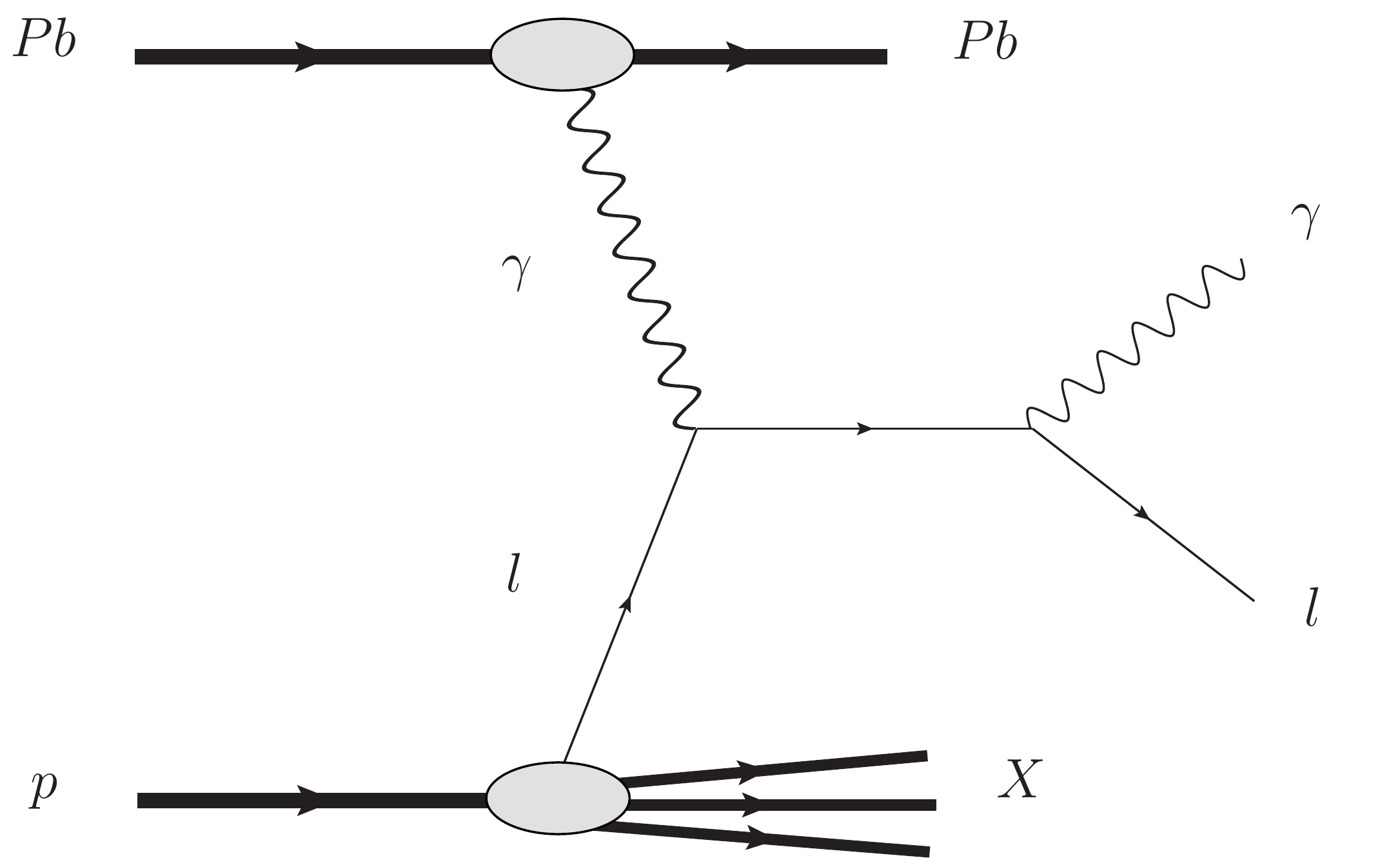}
\caption{Diagrammatic representation of the QED Compton scattering in ultraperipheral $pPb$ collisions, whose cross section is strongly dependent on the leptonic content of the proton.}
\label{Fig:diagram}
\end{figure}

The information about the charged leptons inside the proton can be directly accessed  using a photon as a probe of the hadronic structure.  Such interactions naturally will occur in electron - proton collisions  at the EIC and LHeC \cite{AbdulKhalek:2021gbh,LHeC:2020van}. In contrast, at the LHC, this electromagnetic process is, in principle, very difficult to separate in typical hadronic collisions. The suppression of the photon and lepton PDFs with respect to the quark and gluon PDFs, implies that  the corresponding cross section will be very small, with the associated events being very difficult to separate due to the high pileup present in $pp$ collisions at the LHC. An alternative is to consider ultraperipheral collisions (UPCs), which are characterized by impact parameters larger than the sum of the radii of the incident hadrons \cite{upc}. In this case, the strong interactions between the hadrons are suppressed and the pileups for proton - nucleus and nucleus - nucleus collisions are expected to be small. In addition, if one of the incident hadrons is a nucleus, the associated photon flux is well known and is enhanced by a factor $Z^2$ ($Z$ is the number of protons in the nucleus) with respect to the elastic component of the photon PDF for the proton. These aspects strongly motivate the  study of the QED Compton Scattering in ultraperipheral $pPb$ collisions as a way to constrain the charged lepton PDFs of the proton.  Such a process is represented in Fig. \ref{Fig:diagram}. The main goal of this paper is to estimate, for the first time, the associated cross sections considering different parameterizations for the lepton PDFs currently available in the literature and taking into account the 
 acceptances of the central and forward detectors at the LHC. Moreover, we will explore the possibility of constraining the electron, muon and tau PDFs in the proton, which are predicted to be distinct, providing estimates of the cross sections for different lepton flavours.

\section{Details of the analysis}
The cross section for the QED Compton Scattering in ultraperipheral $pPb$ collisions can be expressed as follows
\begin{eqnarray}
\sigma(Pb p \rightarrow Pb \, \gamma l \, X)  = \int dx_1 dx_2 \, f_{Pb}^{\gamma} (x_1) f_p^l (x_2, \mu^2) \hat{\sigma}(\gamma l \rightarrow \gamma l) \,\,\,,
\label{Eq:cross}  
\end{eqnarray}
where $x_1$ and $x_2$ are the momentum fractions of the nucleus and proton carried by the photon and lepton, respectively,  $\mu$ is a hard scale and $\hat{\sigma}(\gamma l \rightarrow \gamma l) $ is the cross section for the Compton subprocess. One has that $x_{1,2} = [p_T^l \exp(\pm y_l) + p_T^{\gamma} \exp(\pm y_{\gamma})]/\sqrt{s}$, where $p_T^i$ and $y_i$ are the transverse momentum and rapidity of the lepton and photon in the final state, and $\sqrt{s}$ is the center - of - mass energy of the $pPb$ collision. The photon flux for a nuclei $f_{Pb}^{\gamma}$ can be estimated using the Weizs\"acker -- Williams approximation and is given by \cite{upc}
\begin{eqnarray}
f^{\gamma}_A(x) = \frac{\alpha Z^2}{\pi x} \left[ 2 \xi K_0 (\xi) K_1 (\xi) - \xi^2 ( K_1^2 (\xi) - K_0^2 (\xi) ) \right] \,\,,
\end{eqnarray}  
where $K_0$ and $K_1$ are modified Bessel functions and $\xi \equiv x M_A (R_p + R_A)$, with $M_A$ being the projectile mass and $R_p$ ($R_A$)  the proton (nuclear) radius. {As in Ref. \cite{dEnterria:2019jty}, in our analysis we will assume $R_p = 0.7$ fm and $R_A = 7.1$ fm}.   The lepton PDFs in the proton, $f_p^l(x, \mu^2)$, can be estimated  using one of the procedures discussed in the previous Section. In order to estimate the dependence of our predictions on the model assumed for the lepton PDFs, we will calculate the total cross sections and associated differential distributions using the MadGraph 5 NLO \cite{madgraph} and considering the PDF sets derived in Refs. \cite{bertone,buonocore}, which are based on different approaches and predict distinct behaviours for the lepton PDFs at small and large $x$ \cite{buonocore}. As the lepton PDFs derived in Ref. \cite{bertone}  are strongly dependent on the ansatz assumed for the initial condition of the DGLAP evolution and for the underlying photon - PDF sets, we will  consider for completeness of our analysis the  B2, B4, C2 and C4 sets of Ref. \cite{bertone}, which differ by the fact that while the B2 and B4 sets assume that $f_p^l(x, Q_0^2) = 0$,  in the C2 and C4 sets it are calculated from a photon PDF. Moreover, they differ on the underlying PDF sets used in the calculation, which were chosen in Ref. \cite{bertone} as being the NNPDF2.3QED \cite{Ball:2013hta} and MRST2004QED \cite{Martin:2004dh} parametrizations.
  In what follows, the associated predictions will be denoted by  APFEL(NN23NLO)[B2], APFEL(NN23NLO)[C2], APFEL(MRST)[B4] and APFEL(MRST)[C4], respectively. In addition, the cross sections will also be estimated using the lepton PDFs derived in Ref. \cite{buonocore} using the LUX method and the results will be denoted LUXLep. Finally, the factorization scale $\mu$ will be assumed to be equal to the transverse momentum of the lepton in the final state ($p_T^l$).
  
A comment regarding the  state $X$ is in order. As the main contribution for the lepton PDFs comes from the elastic component of the photon PDF \cite{bertone,buonocore}, one can expect that in a large fraction of events  the proton will remain intact, i.e. $X = p$. In this case, the final state of the collision will be characterized by 
two regions devoid of hadronic activity (rapidity gaps) separating the intact very
forward lead nucleus and proton from the central system $\gamma l$. An additional lepton, with opposite charge, that arises from the photon splitting, is also expected to be present in the final state. Such topology can be explored to reduce the possible backgrounds, mainly considering the possibility that the intact proton can be tagged using the ATLAS Forward Proton
detector (AFP) \cite{Adamczyk:2015cjy,Tasevsky:2015xya} and the CMS--Totem Precision Proton Spectrometer
(CT--PPS) \cite{Albrow:2014lrm}. In this paper we will not impose any restriction about the state $X$ and postpone for a future study a detailed analysis of the events characterized by $X = p$.

\begin{table}[t]
\begin{center}
\scalebox{0.92}{
\begin{tabular}{|c|c|c|c|}
\hline 
%{\bf p-p @ $\sqrt{s} = 13$ TeV} & {\bf APFEL(MRST) } & {\bf APFEL(NN23NLO) } & {\bf LUXlep }            \\
% \hline
%ToF and DT                                                        & -    & 0.005,17.65       & -   & -\\\hline 
% Total Cross section (pb)                                       & 0.62  & 0.28   & 1.9    \\\hline 
%p$_{T}(\gamma,e^{+}) > 2.0 ~{\rm GeV}$  &0.62     & 0.28     & 1.9       \\\hline
%$  |\eta|<2.5\:  {\rm(CT-PPS)} $ &0.18 (0.004)  &    0.034 (0.002)    & 0.73 (0.007)     \\\hline
% $ 2.0 < \eta < 4.5 $ & 0.03 & 0.008    &  0.11    \\\hline
  {\bf Pb-p @ $\sqrt{s} = 8.1$ TeV} & {\bf APFEL(MRST)[C4]  } & {\bf APFEL(NN23NLO)[B2]  } & {\bf LUXLep }            \\
Total Cross section (pb)   & e / $\mu$ / $\tau$  &  e/ $\mu$ /  $\tau$  & e/ $\mu$ / $\tau$              \\
 \hline
%Total Cross section (pb) &993.2 / 993.2 / 795.2  & 456.2 / 454.6 / 471.3 & 2100.0 / 1494.0 / 265.6    \\\hline 
$p_{T}^{l,\gamma} \ge 2.0 ~{\rm GeV}$            &     993.2 / 993.2 / 500.0  &456.0 / 456.0 / 308.3     & 2100.0 / 1494.0 / 202.4        \\\hline
$  |\eta_{l,\gamma}|\le 2.5 \:  $ & 387.0 / 387.0 / 168.0  & 124.3 / 124.3 / 77.5        &  1132.0 / 755.5 / 73.6    \\\hline
 $ 2.0 \le \eta_{l,\gamma} \le 4.5 $ & 8.7 / 8.7 / 5.1 & 3.7 / 3.7 / 2.8  & 33.3 / 20.4 / 2.7     \\\hline
\end{tabular}}
\caption{Predictions for the total cross sections of the QED Compton scattering in ultraperipheral $pPb$ collisions at $\sqrt{s} = 8.1$ TeV considering different parametrizations for the lepton PDFs and distinct lepton flavours. The impact of the distinct kinematical cuts is presented. }
\label{tab:cuts_visible_xs}
\end{center}
\end{table}

\section{Results}
In what follows we will present our predictions for the total cross sections and differential distributions for ultraperipheral $pPb$ collisions at $\sqrt{s} = 8.1$ TeV, estimated assuming distinct parameterizations for the lepton PDFs. Our results will be obtained assuming that the ion is right moving.  As emphasized in the Introduction, one of the main advantages of $pPb$ collisions is that the background associated to pileup events is negligible, in contrast to $pp$ collisions. Moreover, background contributions can also be reduced by imposing the presence of a rapidity gap in the final state, associated to the photon exchange, and a cut on the acoplanarity of the photon and lepton jets, such that back - to - back jets are selected. As a consequence, we expect a clean separation of QED Compton scattering in $pPb$ collisions.   In this exploratory study we  will only apply  cuts on the transverse momentum and rapidities of the photon and lepton in the final state. We will require that $p_T^{l}$ and $p_T^{\gamma}$ to be larger than 2.0 GeV and that the  particles are produced in the rapidity range probed by central ($  |\eta_{l,\gamma}|\le 2.5$) or forward ($ 2.0 \le \eta_{l,\gamma} \le 4.5 $)  detectors. In addition, we will calculate the cross sections for the different lepton flavours. Our results are presented in Table \ref{tab:cuts_visible_xs} for the APFEL(NN23NLO)[B2], APFEL(MRST)[C4] and LUXLep sets.  One has that the predictions are strongly dependent on the lepton PDFs assumed in the calculation, with the LUXLep set predicting the higher (lower) values for electron and muon (tau) production. Moreover, the LUXLep set predicts distinct cross sections for the different lepton flavours, while those obtained using APFEL imply identical results for electron and muon production. Such result is associated to the fact that our predictions are dominated by the contribution of lepton PDFs at small values of $x$, where the  APFEL(MRST)[C4] predictions for the electron and muon PDFs are very similar (See Fig. 2 in Ref. \cite{bertone}). One has that the cut on the rapidities of the photon and lepton in the final state imply a reduction by a factor $\approx 3 \, (100)$ for a central (forward) selection. Regarding the results for the cross sections, we predict values of the order of pb, which implies a non-negligible number of events per year in the future runs of the LHC, especially for the central selection and electron production.

\begin{figure}[t]
\includegraphics[scale=0.42]{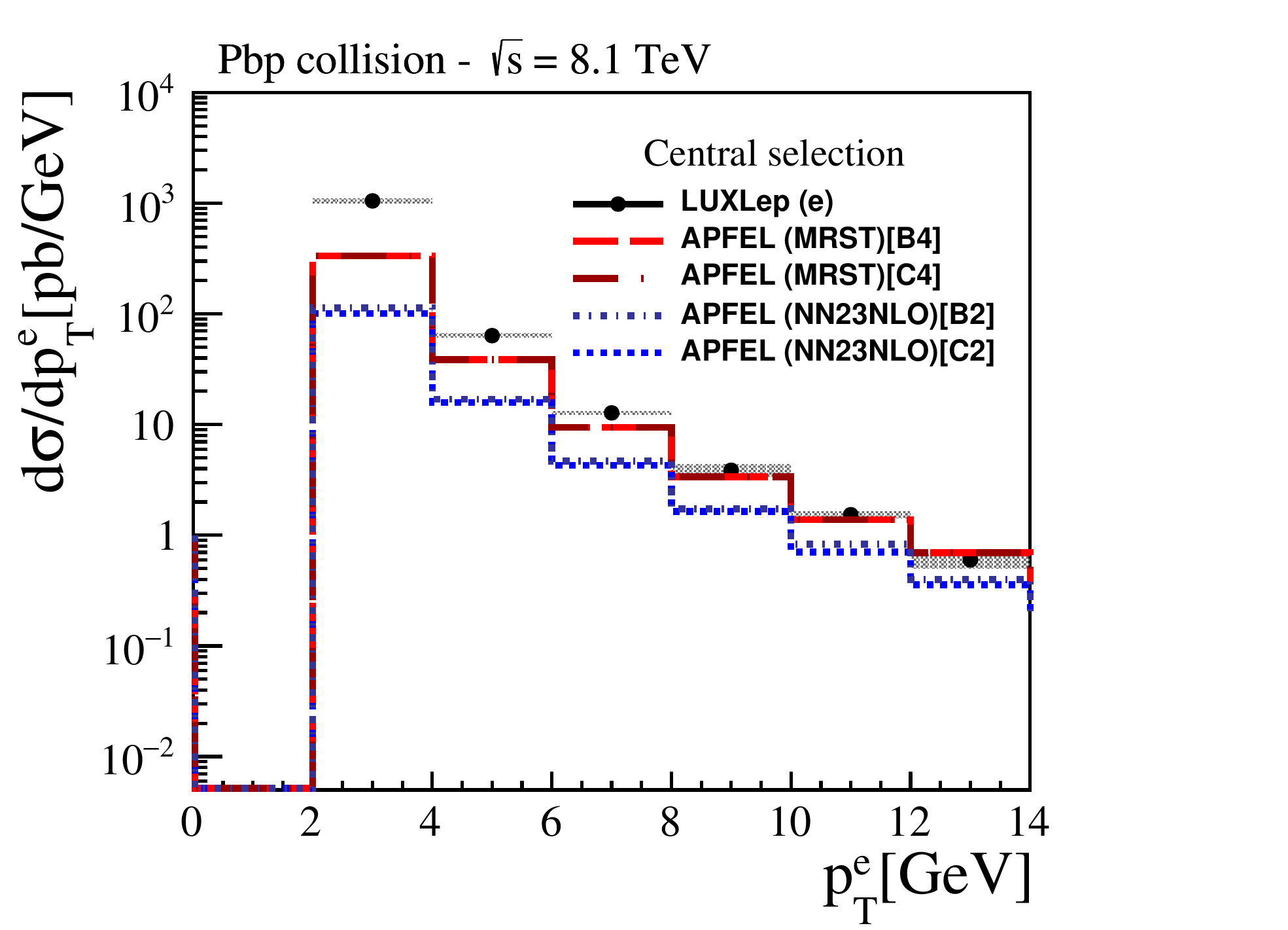} 
\includegraphics[scale=0.42]{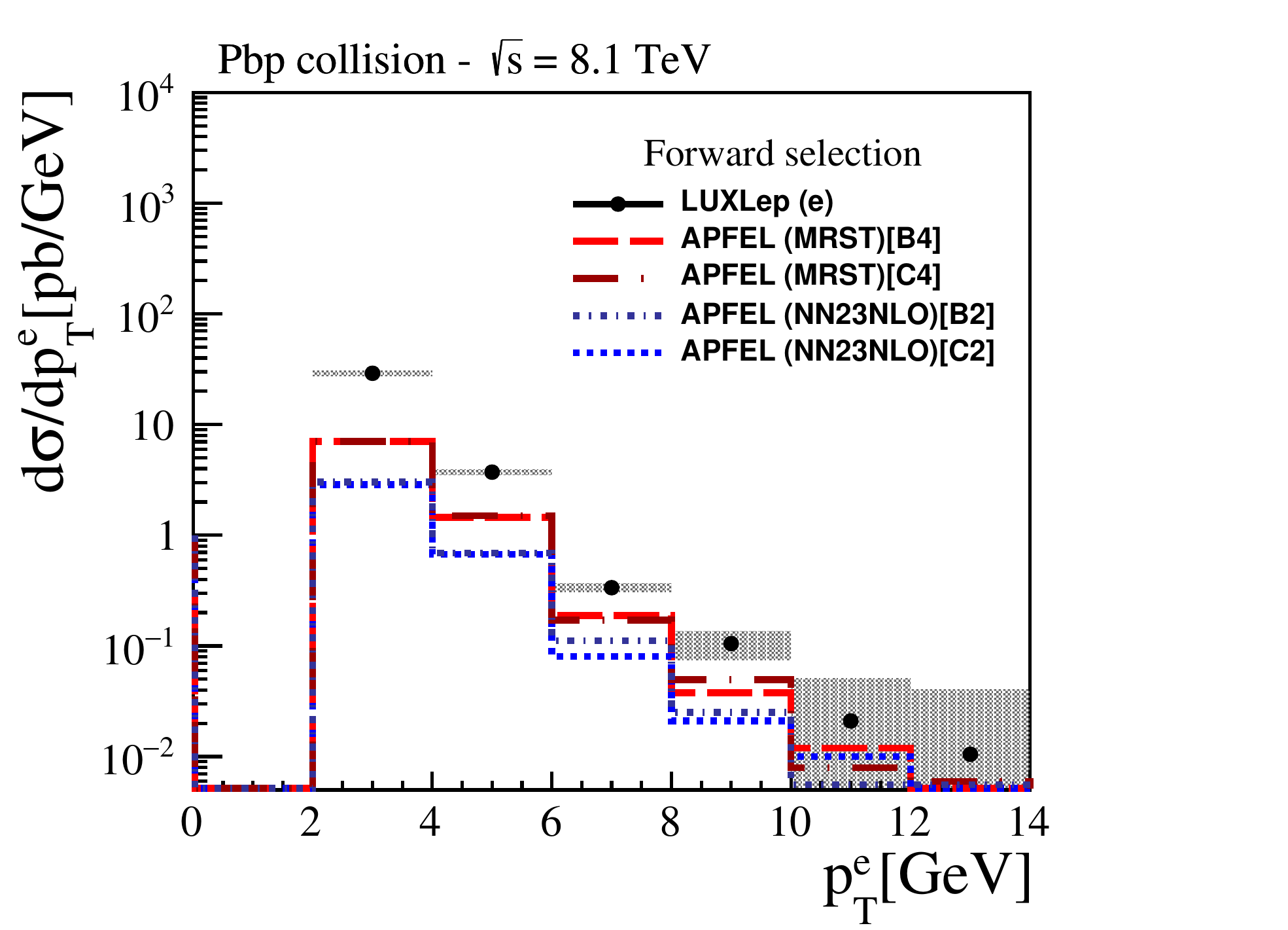} 
\includegraphics[scale=0.42]{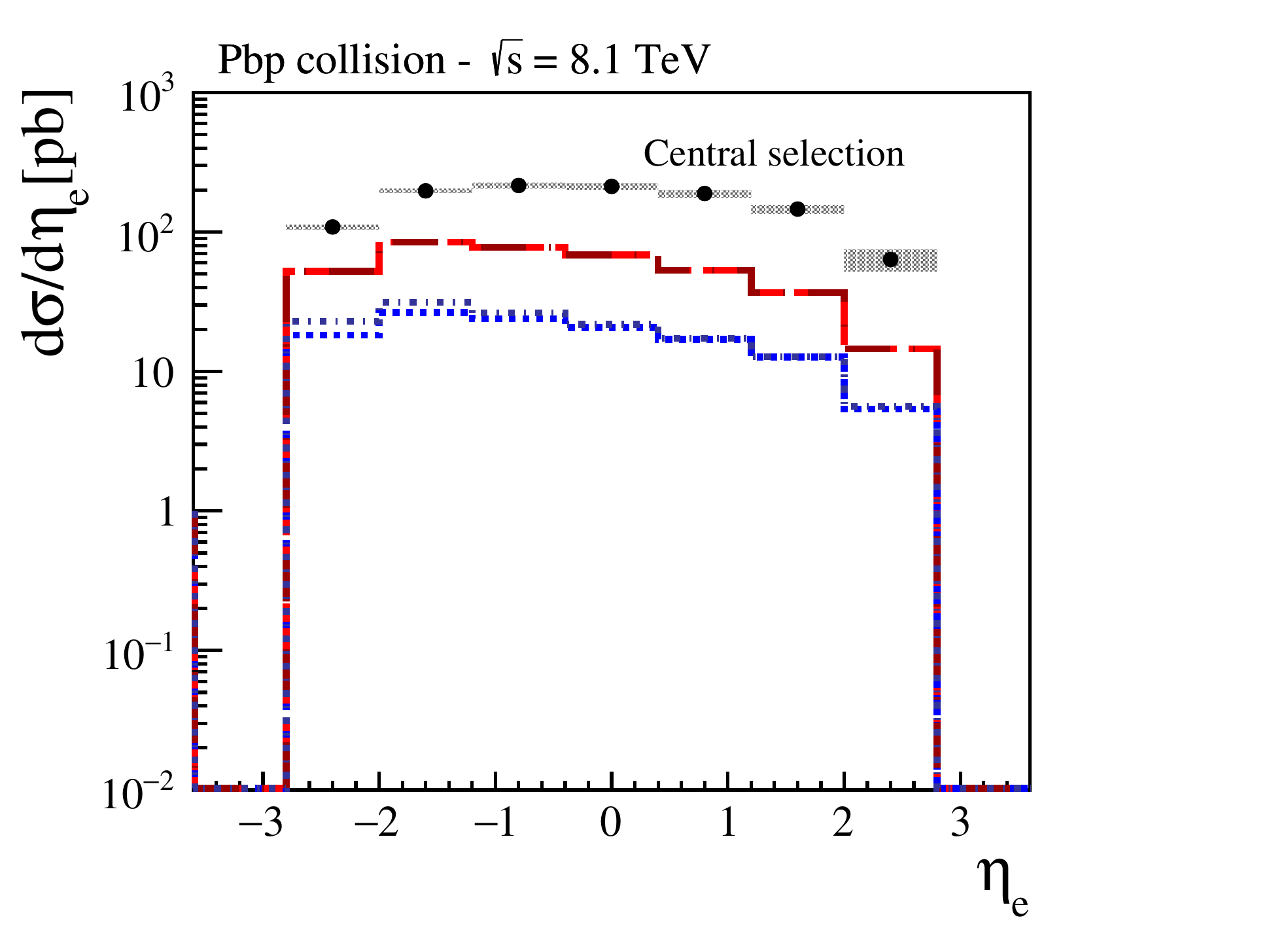} 
\includegraphics[scale=0.42]{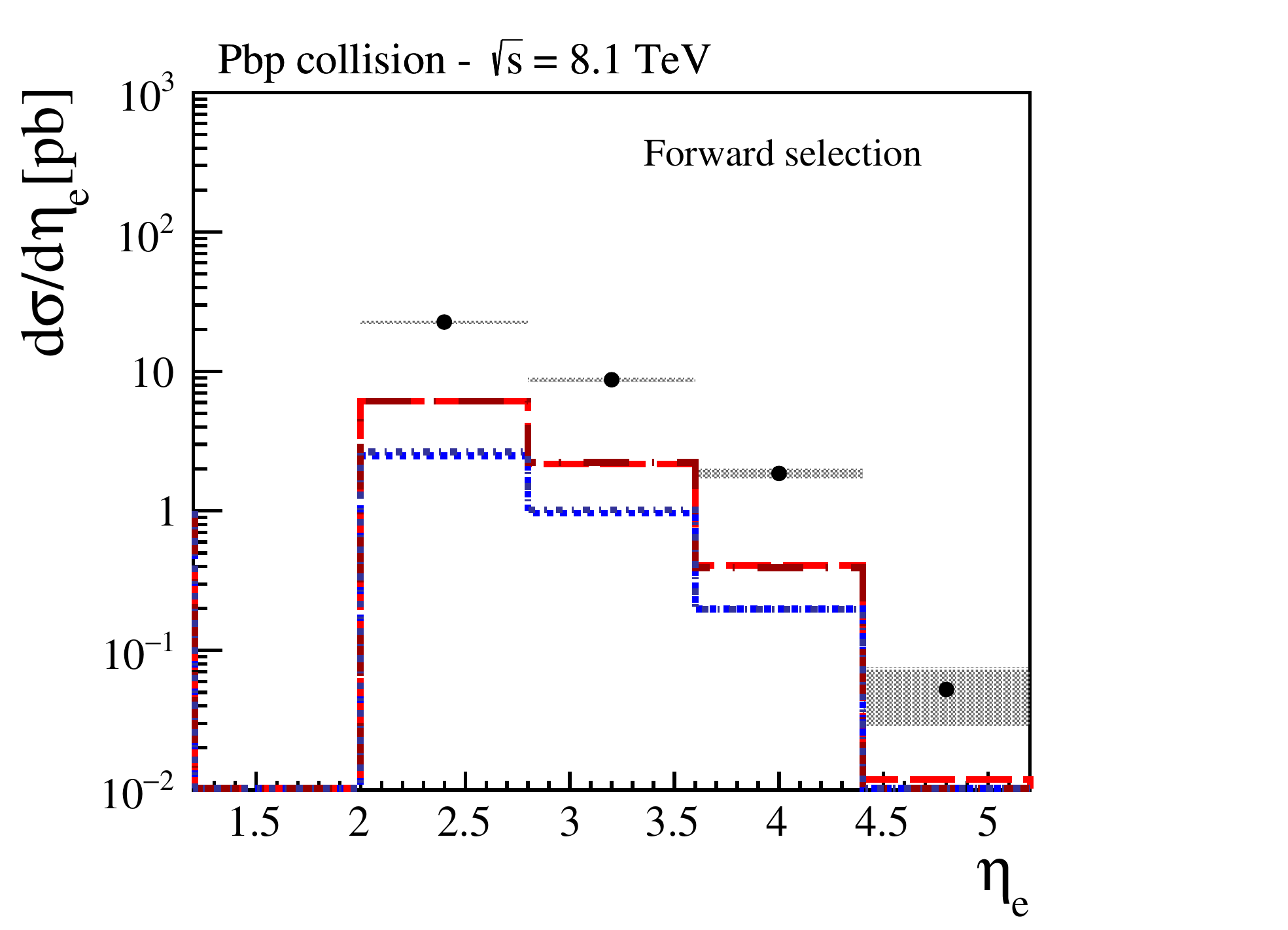}
\caption{Transverse momentum (upper panels) and rapidity (lower panels) distributions for the electron production in the QED Compton scattering in ultraperipheral $pPb$ collisions at $\sqrt{s} = 8.1$ TeV, calculated considering different leptons PDFs and the central (left panels) and forward (right panels) selections.}
\label{Fig:Pbp_8TeV_difpdfs}
\end{figure} 

The predictions for the transverse momentum and rapidity distributions are presented in Fig. \ref{Fig:Pbp_8TeV_difpdfs} considering the electron production, distinct parametrizations for the lepton PDFs and the central and forward selections. For completeness the LUXLep uncertainties are taken into account and represented by the shadowed band. One has that the APFEL predictions are dependent on the assumption for the initial condition, as can be verified comparing the APFEL(NN23NLO)[B2] and APFEL(NN23NLO)[C2] results. On the other hand, the results obtained using distint parametrizations in the derivation of the lepton PDFs are almost identical, as can be observed comparing e.g. the results denoted APFEL(NN23NLO)[C2] and APFEL(MRST)[C4] in the figure.    As expected from the analysis of the total cross sections, the APFEL(MRST)[C4], APFEL(NN23NLO)[B2] and LUXLep parametrizations   imply different predictions for the distributions. In particular, the rapidity distribution is strongly sensitive to the PDF set, which is directly associated to the distinct behaviours of the lepton PDFs  for small and large $x$. One also has verified that the invariant mass distribution of the final state (not shown) is sensitive to the lepton PDF set assumed in the calculation, which implies that this distribution is also an alternative to constrain the leptonic densities in the proton.

\begin{figure}[t]
\includegraphics[scale=0.42]{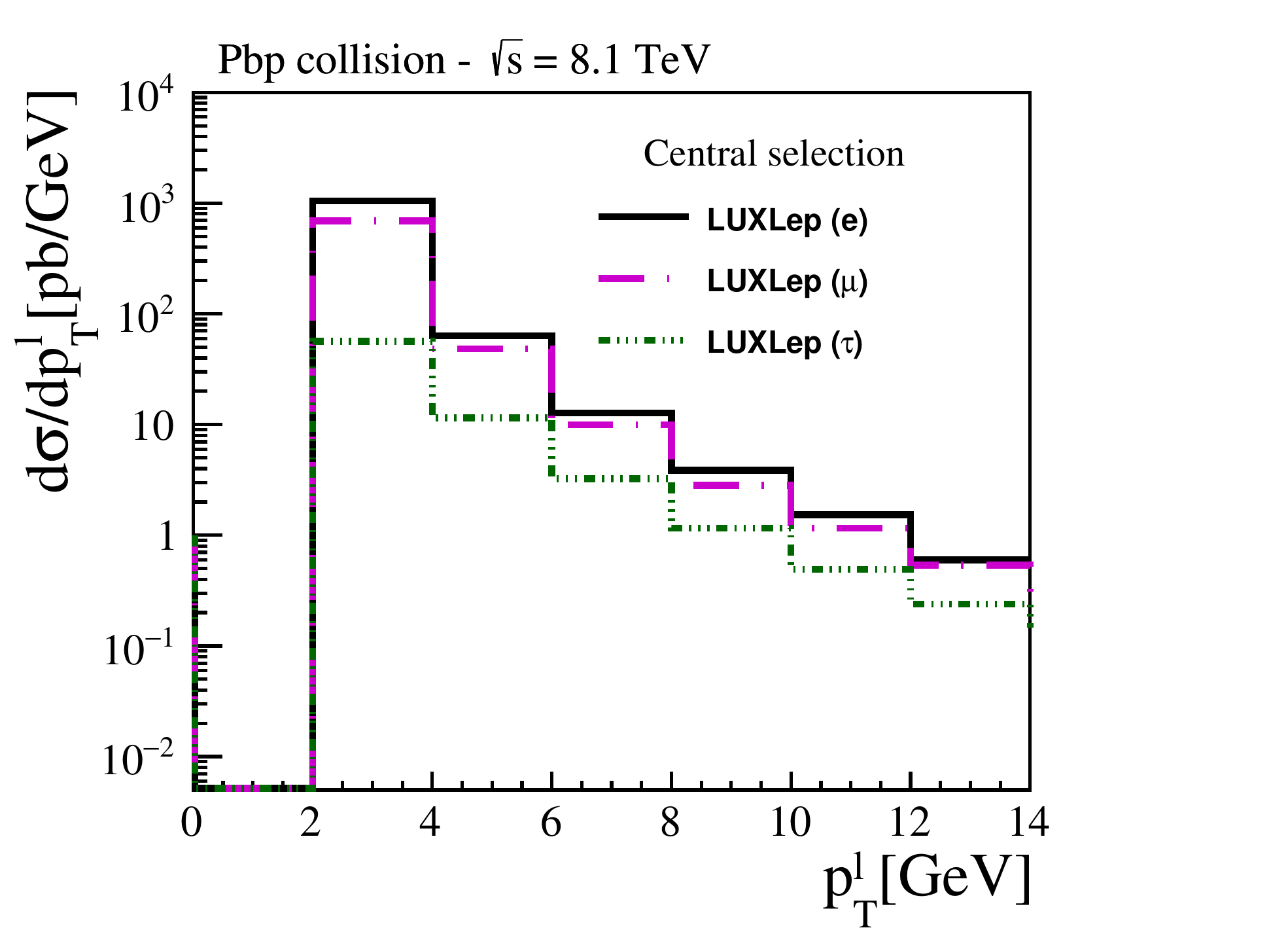} 
\includegraphics[scale=0.42]{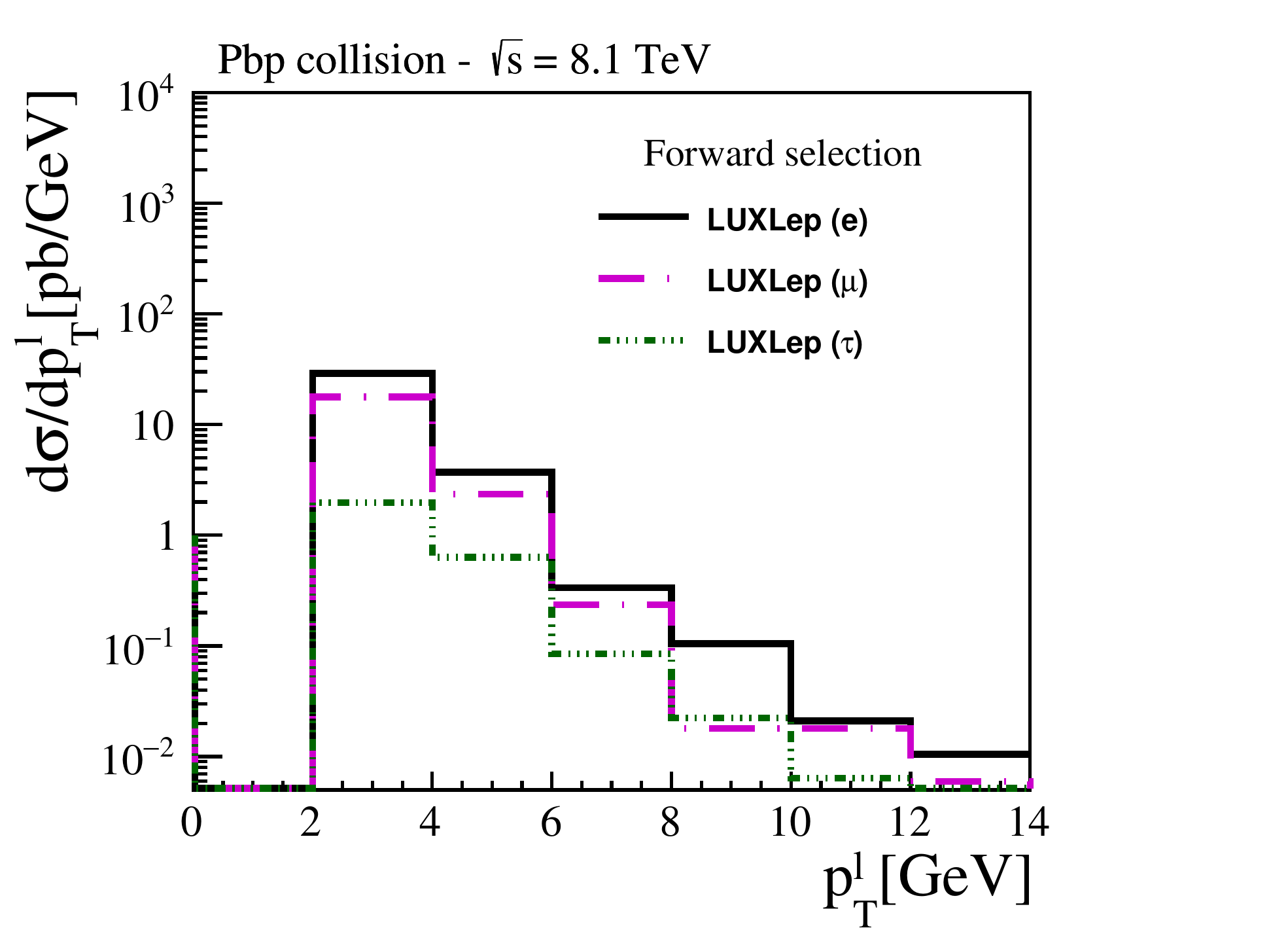} 
\includegraphics[scale=0.42]{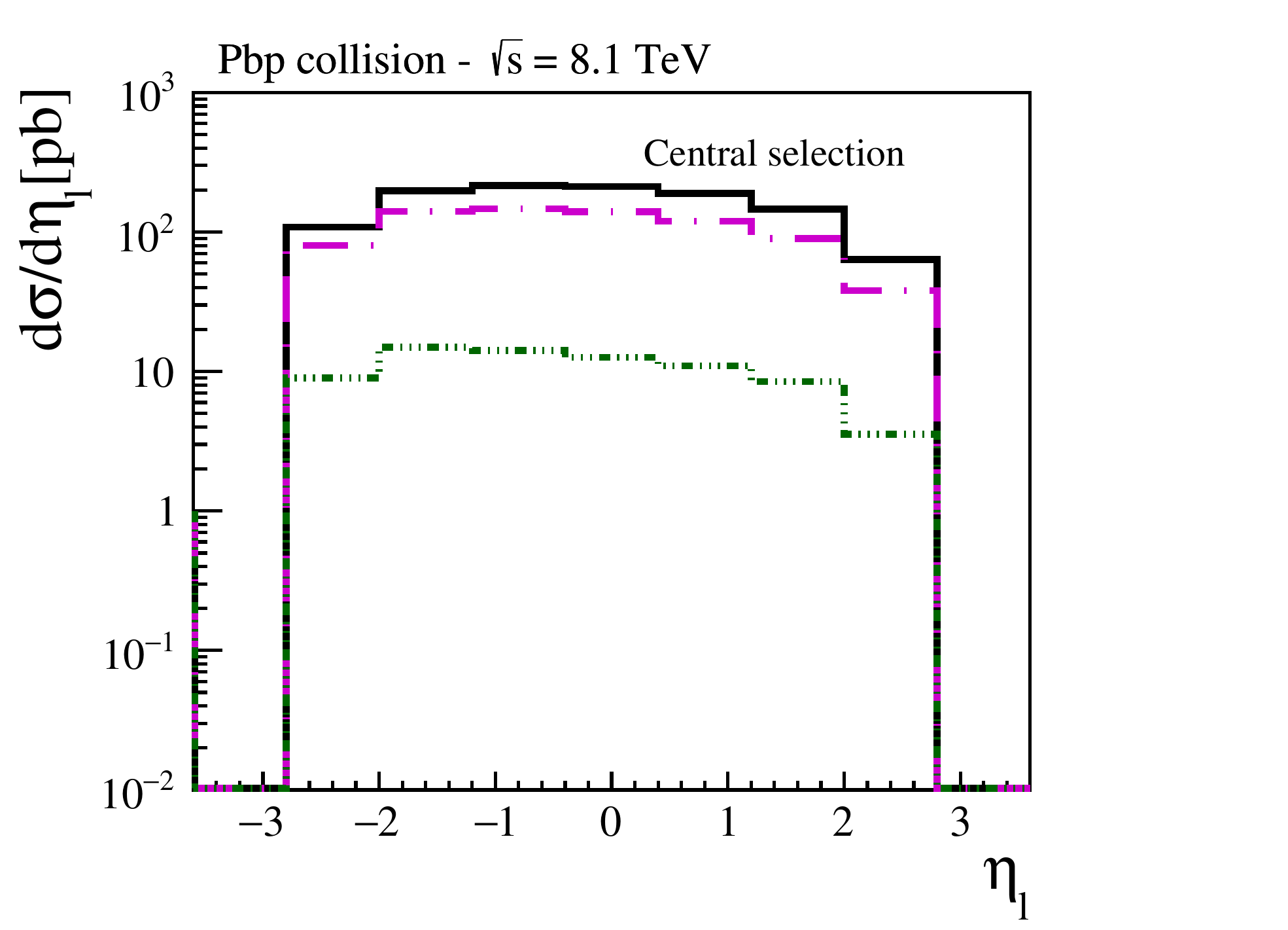} 
\includegraphics[scale=0.42]{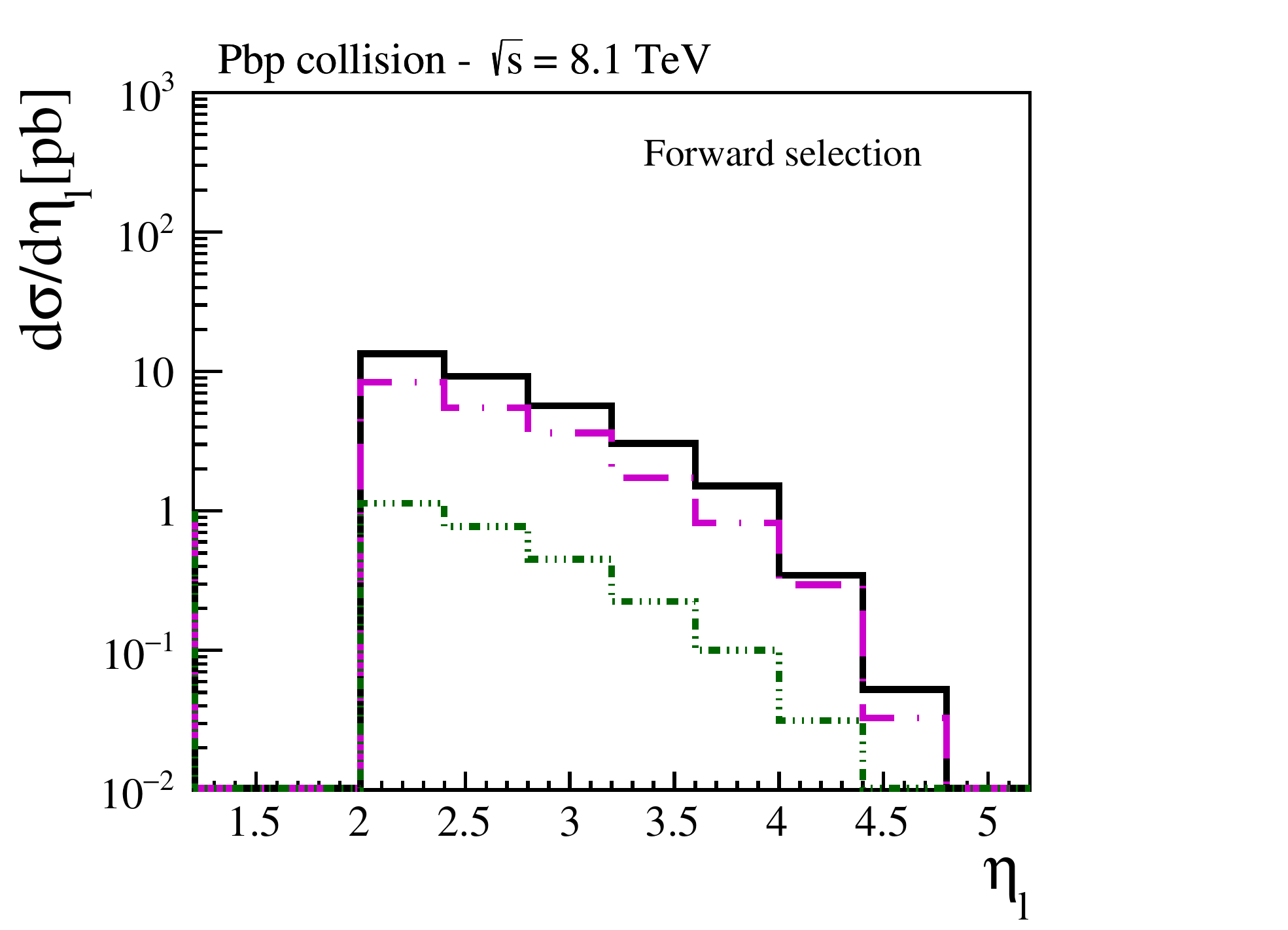}
\caption{Transverse momentum (upper panels) and rapidity (lower panels) distributions for the  QED Compton scattering in ultraperipheral $pPb$ collisions at $\sqrt{s} = 8.1$ TeV calculated considering the LUXlep parametrization for different lepton flavours and the central (left panels) and forward (right panels) selections. }
\label{Fig:Pbp_8TeV_lux}
\end{figure}

In Fig. \ref{Fig:Pbp_8TeV_lux} we present our predictions for the transverse momentum and rapidity distributions of the QED Compton scattering in ultraperipheral $pPb$ collisions at $\sqrt{s} = 8.1$ TeV considering the LUXLep parametrization for the different lepton flavours and  the central  and forward  selections. One has that the predictions are sensitive to the lepton flavour, in agreement with the results presented in Table \ref{tab:cuts_visible_xs}. The difference between the predictions for the  production of the distinct flavours is non - negligible, which implies that the associated lepton PDFs can, in principle, be separately constrained in a future analysis of this process.

Before summarizing our main results and conclusions, it is important to comment about the possibility of studying the QED Compton scattering in $pp$ collisions at the LHC. One has verified that the resulting predictions for electron production, obtained using the LUXLep set and $\sqrt{s} = 13$ TeV, are of the order of 0.7 pb for the central selection and similar behaviours for the distributions are predicted. As discussed above, the separation of these events is a hard task due to the large pileup in current and future runs of the LHC. The situation can be improved by imposing the exclusivity criteria, i.e. that the protons remain intact in the final state and are tagged for the forward detectors AFP and CT - PPS (For a more detailed discussion, see e.g. Refs. \cite{top1,top2}). However, for this tagging to be possible, the invariant mass of the final state should be larger than 200 GeV. Imposing this cut on our results, we have verified that the associated cross section is reduced in two orders of magnitude. Therefore, we predict values of the order of  few fb, which must be considered an upper bound since in some of the events considered in our analysis the proton that emits the lepton is expected to break up. These results indicate that the measurement of QED Compton scattering in $pp$ collisions in the high pileup runs of the LHC will not be feasible. An alternative is to perform the analysis of this process during a special run with low pileup, where the events can be separated by observing the rapidity gaps in the final state and lower values of the invariant mass are accessible.

\section{Summary}
The understanding of the structure of the proton is one of the challenging problems in Particle Physics. In addition to the valence quarks needed to describe its quantum numbers, the quantum dynamical evolution of the proton, implies the presence of a large number of gluons and sea quarks, as well as photons and leptons as its constituents, which have an associated parton distribution function. The precise determination of these PDFs is fundamental to describe the standard model processes at the LHC  and the searching for New Physics. In particular, the presence of  leptons inside the proton makes the LHC also a lepton - lepton and lepton - quark collider, which can explore new production mechanisms. In recent years, distinct groups have proposed different parameterizations for the lepton PDFs, which differ in normalization and behaviors for   small and large momentum fractions,  implying that the current predictions for the lepton - induced processes  are still affected by a reasonable theoretical uncertainty. In this paper, we have proposed, for the first time, to constrain the lepton PDFs through the analysis of the QED Compton scattering in ultraperipheral $pPb$ collisions at the LHC. In this process, the lead ion is the source of a large number of photons that illuminate the  leptonic  constituents of the proton. We have estimated the total cross sections and associated distributions considering different models for the lepton PDFs and demonstrated that the predictions are sensitive to the parameterization considered as well as to the lepton flavour. In addition, our results indicated that a future analysis of the process is feasible. Finally, the possibility of study of the QED Compton scattering in $pp$ collisions was also discussed and alternatives to measure this process were pointed out. The results presented in this paper strongly motivate a more detailed analysis that we plan to do in a forthcoming study.

\begin{acknowledgments}
The authors thank Valerio Bertone for providing the APFEL lepton PDFs.
This work was partially supported by CNPq, CAPES, FAPERGS and  INCT-FNA (Process No. 464898/2014-5).  V.P.G. was also partially supported by the CAS President's International Fellowship Initiative (Grant No.  2021VMA0019). D.E.M. was supported by CNPq (Grant no. 152473/2022-0) and by the Henryk Niewodniczanski Institute of Nuclear Physics Polish Academy of Sciences (grant no. UMO-2021/43/P/ST2/02279).  
\end{acknowledgments}

\end{document}